\documentclass[pra,aps,showpacs,onecolumn,twoside,superscriptaddress]{revtex4-1}

\usepackage{amsmath,amsfonts,amssymb,caption,color,epsfig,graphics,graphicx,hyperref,latexsym,mathrsfs,revsymb,theorem,url,verbatim,epstopdf}

\usepackage{amssymb}
\usepackage{color}
\usepackage{amsmath,epic,curves,amscd}
\usepackage[english]{babel}
\usepackage{graphicx}
\usepackage{comment}
\usepackage{appendix}
\usepackage{mathdots}

\newtheorem{definition}{Definition}
\newtheorem{proposition}[definition]{Proposition}
\newtheorem{lemma}[definition]{Lemma}

\newtheorem{theorem}[definition]{Theorem}
\newtheorem{corollary}[definition]{Corollary}
\newtheorem{conjecture}[definition]{Conjecture}

\newtheorem{remark}[definition]{Remark}
\newtheorem{example}[definition]{Example}
\newtheorem{question}[definition]{Question}

\def\squareforqed{\hbox{\rlap{$\sqcap$}$\sqcup$}}
\def\qed{\ifmmode\squareforqed\else{\unskip\nobreak\hfil
\penalty50\hskip1em\null\nobreak\hfil\squareforqed
\parfillskip=0pt\finalhyphendemerits=0\endgraf}\fi}
\def\endenv{\ifmmode\;\else{\unskip\nobreak\hfil
\penalty50\hskip1em\null\nobreak\hfil\;
\parfillskip=0pt\finalhyphendemerits=0\endgraf}\fi}
\newenvironment{proof}{\noindent \textbf{{Proof.~} }}{\qed}
\def\Dbar{\leavevmode\lower.6ex\hbox to 0pt
{\hskip-.23ex\accent"16\hss}D}
\makeatletter
\def\url@leostyle{%
  \@ifundefined{selectfont}{\def\UrlFont{\sf}}{\def\UrlFont{\small\ttfamily}}}
\makeatother
\def\bcj{\begin{conjecture}}
\def\ecj{\end{conjecture}}
\def\bcr{\begin{corollary}}
\def\ecr{\end{corollary}}
\def\bd{\begin{definition}}
\def\ed{\end{definition}}
\def\bea{\begin{eqnarray}}
\def\eea{\end{eqnarray}}
\def\bem{\begin{enumerate}}
\def\eem{\end{enumerate}}
\def\bex{\begin{example}}
\def\eex{\end{example}}
\def\bim{\begin{itemize}}
\def\eim{\end{itemize}}
\def\bl{\begin{lemma}}
\def\el{\end{lemma}}
\def\bma{\begin{bmatrix}}
\def\ema{\end{bmatrix}}
\def\bpf{\begin{proof}}
\def\epf{\end{proof}}
\def\bpp{\begin{proposition}}
\def\epp{\end{proposition}}
\def\bqu{\begin{question}}
\def\equ{\end{question}}
\def\br{\begin{remark}}
\def\er{\end{remark}}
\def\bt{\begin{theorem}}
\def\et{\end{theorem}}

\def\btb{\begin{tabular}}
\def\etb{\end{tabular}}

\newcommand{\nc}{\newcommand}

\def\a{\alpha}
\def\b{\beta}
\def\g{\gamma}
\def\d{\delta}
\def\e{\epsilon}

\def\z{\zeta}

\def\t{\theta}
\def\i{\iota}
\def\k{\kappa}

\def\m{\mu}

\def\x{\xi}

\def\r{\rho}
\def\s{\sigma}

\def\u{\upsilon}

\def\c{\chi}

\def\G{\Gamma}

\def\T{\Theta}

 \nc{\bbA}{\mathbb{A}} \nc{\bbB}{\mathbb{B}} \nc{\bbC}{\mathbb{C}}
 \nc{\bbD}{\mathbb{D}} \nc{\bbE}{\mathbb{E}} \nc{\bbF}{\mathbb{F}}
 \nc{\bbG}{\mathbb{G}} \nc{\bbH}{\mathbb{H}} \nc{\bbI}{\mathbb{I}}
 \nc{\bbJ}{\mathbb{J}} \nc{\bbK}{\mathbb{K}} \nc{\bbL}{\mathbb{L}}
 \nc{\bbM}{\mathbb{M}} \nc{\bbN}{\mathbb{N}} \nc{\bbO}{\mathbb{O}}
 \nc{\bbP}{\mathbb{P}} \nc{\bbQ}{\mathbb{Q}} \nc{\bbR}{\mathbb{R}}
 \nc{\bbS}{\mathbb{S}} \nc{\bbT}{\mathbb{T}} \nc{\bbU}{\mathbb{U}}
 \nc{\bbV}{\mathbb{V}} \nc{\bbW}{\mathbb{W}} \nc{\bbX}{\mathbb{X}}
 \nc{\bbZ}{\mathbb{Z}}

 \nc{\bA}{{\bf A}} \nc{\bB}{{\bf B}} \nc{\bC}{{\bf C}}
 \nc{\bD}{{\bf D}} \nc{\bE}{{\bf E}} \nc{\bF}{{\bf F}}
 \nc{\bG}{{\bf G}} \nc{\bH}{{\bf H}} \nc{\bI}{{\bf I}}
 \nc{\bJ}{{\bf J}} \nc{\bK}{{\bf K}} \nc{\bL}{{\bf L}}
 \nc{\bM}{{\bf M}} \nc{\bN}{{\bf N}} \nc{\bO}{{\bf O}}
 \nc{\bP}{{\bf P}} \nc{\bQ}{{\bf Q}} \nc{\bR}{{\bf R}}
 \nc{\bS}{{\bf S}} \nc{\bT}{{\bf T}} \nc{\bU}{{\bf U}}
 \nc{\bV}{{\bf V}} \nc{\bW}{{\bf W}} \nc{\bX}{{\bf X}}
 \nc{\bZ}{{\bf Z}}

\nc{\cA}{{\cal A}} \nc{\cB}{{\cal B}} \nc{\cC}{{\cal C}}
\nc{\cD}{{\cal D}} \nc{\cE}{{\cal E}} \nc{\cF}{{\cal F}}
\nc{\cG}{{\cal G}} \nc{\cH}{{\cal H}} \nc{\cI}{{\cal I}}
\nc{\cJ}{{\cal J}} \nc{\cK}{{\cal K}} \nc{\cL}{{\cal L}}
\nc{\cM}{{\cal M}} \nc{\cN}{{\cal N}} \nc{\cO}{{\cal O}}
\nc{\cP}{{\cal P}} \nc{\cQ}{{\cal Q}} \nc{\cR}{{\cal R}}
\nc{\cS}{{\cal S}} \nc{\cT}{{\cal T}} \nc{\cU}{{\cal U}}
\nc{\cV}{{\cal V}} \nc{\cW}{{\cal W}} \nc{\cX}{{\cal X}}
\nc{\cY}{{\cal Y}}
\nc{\cZ}{{\cal Z}}

\nc{\hA}{{\hat{A}}} \nc{\hB}{{\hat{B}}} \nc{\hC}{{\hat{C}}}
\nc{\hD}{{\hat{D}}} \nc{\hE}{{\hat{E}}} \nc{\hF}{{\hat{F}}}
\nc{\hG}{{\hat{G}}} \nc{\hH}{{\hat{H}}} \nc{\hI}{{\hat{I}}}
\nc{\hJ}{{\hat{J}}} \nc{\hK}{{\hat{K}}} \nc{\hL}{{\hat{L}}}
\nc{\hM}{{\hat{M}}} \nc{\hN}{{\hat{N}}} \nc{\hO}{{\hat{O}}}
\nc{\hP}{{\hat{P}}} \nc{\hR}{{\hat{R}}} \nc{\hS}{{\hat{S}}}
\nc{\hT}{{\hat{T}}} \nc{\hU}{{\hat{U}}} \nc{\hV}{{\hat{V}}}
\nc{\hW}{{\hat{W}}} \nc{\hX}{{\hat{X}}} \nc{\hZ}{{\hat{Z}}}

\nc{\hn}{{\hat{n}}}

\def\dim{\mathop{\rm Dim}}

\def\max{\mathop{\rm max}}
\def\min{\mathop{\rm min}}

\def\rank{\mathop{\rm rank}}

\def\w{\mathop{\rm W}}

\def\ox{\otimes}

\def\LRa{\Longrightarrow}

\newcommand{\ket}[1]{|#1\rangle}

\newcommand{\jpa}{J. Phys. A}

\def\Dbar{\leavevmode\lower.6ex\hbox to 0pt
{\hskip-.23ex\accent"16\hss}D}

\begin{document}

\title{Multiqubit UPB: The method of formally orthogonal 
matrices}

\author{Lin Chen}
\email{linchen@buaa.edu.cn (corresponding author)}
\affiliation{School of Mathematics and Systems Science, Beihang University, Beijing 100191, China}
\affiliation{International Research Institute for Multidisciplinary Science, Beihang University, Beijing 100191, China}

\def\Dbar{\leavevmode\lower.6ex\hbox to 0pt
{\hskip-.23ex\accent"16\hss}D}
\author {{ Dragomir {\v{Z} \Dbar}okovi{\'c}}}
\email{djokovic@uwaterloo.ca}
\affiliation{Department of Pure Mathematics and Institute for
Quantum Computing, University of Waterloo, Waterloo, Ontario, N2L
3G1, Canada} 

\date{\today}


\begin{abstract}
We use formal matrices whose entries we view as vector variables taking unit vectors values in one-qubit Hilbert spaces of a 
multiqubit quantum system. We construct many unextendible product bases (UPBs) of new sizes in such systems and provide 
a new construction of UPBs of $n$ qubits of cardinality $n+1$ 
when $n\equiv 3 \pmod{4}$. We also give a new method of 
constructing multiqubit entangled states with all partial transposes positive.
\end{abstract}

\maketitle

\tableofcontents

\section{Introduction}

We introduced in \cite{cd17} a formal matrix approach in order 
to study the orthogonal product bases (OPB) in multiqubit 
systems. We applied this method to obtain a coarse classification of OPBs of four qubits. We warn the reader that 
this is not the classification under local unitary transformations and qubit permutations.

In the present paper we extend this formal approach in order to 
study the unextendible product bases (UPB) in multiqubit systems. We adapted many definitions from \cite{cd17} and introduced some new ones. The entries of our formal matrices, 
$X$, are vector variables which take unit vector values (up to the phase factor) in one of the single qubit Hilbert spaces. 
The column $j$ of the matrix corresponds to the $j$th qubit.

The powerful tool of so called orthogonality graphs is naturally 
embedded into our formal matrix framework. Each vector variable, say $x$, has a companion, its perpendicular, which we denote by $x'$. If $x$ takes a value $\ket{a}$ then $x'$ takes as its value the unique unit vector $\ket{a}^\perp$ orthogonal to $\ket{a}$, in the Hilbert space of the same qubit. In this way the rows of the formal matrix $X$ give rise to product vectors. The rows of $X$ are the vertices of the orthogonality graphs. 
We obtain the orthogonality graph of the $j$th qubit 
by joining the vertices $i$ and $k$ if and only if the matrix 
entries in the positions $(i,j)$ and $(k,j)$ are the perpendiculars of each other. The rows $i$ and $k$ are 
orthogonal if such $j$ exists. The matrix is orthogonal if any 
two of its rows are orthogonal. An orthogonal matrix $X$ is an unextendible orthogonal matrix (UOM) if there is no row orthogonal to all of the rows of $X$. 
(See the next section for precise definitions.)
Each UOM $X$ gives an infinite family of UPBs, which we denote by $\cF_X^\#$, having the same orthogonality graphs as $X$.

The formal method makes it possible to simplify the proofs of
many known facts, to generalize some of the known results, 
and is conducive to making new constructions and conjectures. 
For instance, by using Lemma \ref{le:uom-1} in 
Sec. \ref{sec:pptes} we construct a new class of PPTES 
(entangled states, all of whose partial transposes are positive semidefinite) which we call secondary PPTES. 

One of the main open problems in the study of the UPBs in
$n$-qubit quantum systems is to determine the set $\T_n$ of
possible sizes of UPBs for fixed $n$. In the language of formal matrices, this set is the set of integers $m$ for which there exist a UOM of size  $m\times n$. Many results about the sets $\T_n$ are known. In particular, the smallest integer $\t_n$ of $\T_n$ is known \cite{johnston13,johnston14}. 
On the other hand, many problems were left open in these papers.
One of them was whether $2^n-5\in\T_n$, to which a negative answer was obtained recently \cite{cd17}. 
For $n\le7$, the question whether $m\in\T_n$ was left open also when $(m,n)$ is one of the following pairs: 
$(11,5)$, $(10,6)$, $(11,6)$, $(13,6)$, $(10,7)$, $(11,7)$, $(13,7)$, $(14,7)$, $(15,7)$, $(19,7)$. 
We have constructed many UOMs of new sizes $m\times n$, in 
particular for all pairs listed above except $(10,7)$ and $(11,7)$ which remain open. As a consequence, the sets $\T_n$ 
are now known exactly for $n\le 6$ and for $n=8$. Previously 
they were known only for $n\le4$.

The UPBs provide nice examples of PPTES. It is well known that 
the projector $\r$ onto the subspace of $\cH$ orthogonal to a UPB, say $\cU$, is a (non-normalized) PPTES. The range of $\r$ contains no product vectors. We modify this construction as follows. Let us drop from $\cU$ one of the product vectors and denote by $\s$ the projector onto the subspace orthogonal to the remaining product vectors of $\cU$. We show that the range of 
$\s$ contains only finitely many product vectors (up to scalar 
mutiples). In many cases (but not always) $\s$ is a (non-normalized) PPTES. Its  range contains at least one product vector, and so these new PPTES are essentially different from the PPTES $\r$ mentioned above. Our Proposition \ref{pp:kon} in Sec. \ref{app:B} shows that this construction of PPTES is also applicable to UPBs in arbitrary finite-dimensional quantum systems.

The rest of the paper is organized as follows. In Sec. \ref{sec:pre} we generalize our formal matrix approach from \cite{cd17} in order to be able to apply it to the more general case of arbitrary UPBs (in multiqubit systems). We describe the basic concepts and give the definitions used in this paper, such as orthogonal product bases, vector variables, orthogonal matrices, and equivalence of matrices. 
In Sec. \ref{sec:extensions} we introduce the concepts of the 
evaluations and extensions of orthogonal matrices, define the UOMs and describe a method of constructing larger orthogonal matrices or UOMs from the smaller ones, see Proposition \ref{pp:konstr}. 
In Sec. \ref{sec:facts}, Theorem \ref{thm:upb}, 
we review some known facts on the existence of UPBs and 
rephrase them in terms of UOMs. The UOMs of new sizes that we 
have constructed are listed in Table \ref{tab:sizes}. 
It is important to have a good test for checking whether 
two UOMs of the same size are equivalent. The test that we 
used is based on Lemma \ref{le:eq-test}.
In Sec. \ref{sec:newconstr} we describe a new construction of 
orthogonal matrices which can be used to produce new UOMs.
For instance, for any $n\equiv 3 \pmod{4}$ we can construct new 
UOMs of size $(n+1)\times n$. In particular, for $n=7$ our 
construction gives four non-equivalent UOMs. (There are 
only 7 equivalence classes of UOMs of that size.)
In Sec. \ref{sec:pptes} we describe a new type of PPTES 
associated to UPBs to which we refer as the secondary PPTES. 
In the case of four qubits, we list in Table \ref{tab:secondary}
all pairs $(\rank\r,s)$ where $\r$ is a secondary PPTES 
and $s$ is the number of product vectors in the range of $\r$.
In Sec. \ref{sec:maximal} we define a partial order in $\cM(m,n)$ and $\cO(m,n)$, and use it to deduce a partial order 
on the set of equivalence classes of orthogonal matrices 
as well as the equivalence classes of UOMs.

\section{Basic concepts and definitions}
\label{sec:pre}

Let $\cH=\cH_1\ox\cdots\ox\cH_n$ be the Hilbert space representing a quantum system $A_1,\cdots,A_n$ consisting of $n$ qubits. 
Each $\cH_j$ is a 2-dimensional Hilbert space. We fix an 
orthonormal basis $\ket{0}_j,\ket{1}_j$ of $\cH_j$. Usually, 
the subscript $j$ will be suppressed. 
We say that a vector $\ket{v}\in\cH$ is a unit vector 
if $\|v\|=1$. For any nonzero vector $\ket{v_j}\in\cH_j$ 
we denote by $[v_j]$ the 1-dimensional subspace of 
$\cH_j$ spanned by this vector. As a rule, we shall not 
distinguish two unit vectors which differ only in the 
phase. By using this convention, we can say that for 
any unit vector $\ket{v_j}\in\cH_j$ there exists a unique 
unit vector $\ket{v_j}^\perp\in\cH_j$ which is perpendicular 
to $\ket{v_j}$. 

A product vector is a nonzero vector 
$\ket{x}=\ket{x_1}\ox\cdots\ox\ket{x_n}$, which will be 
written also as $\ket{x}=\ket{x_1,\ldots,x_n}$. 
If $\|x\|=1$ we shall assume (as we may) that 
each $\|x_j\|=1$. 
Two product vectors $\ket{x}=\ket{x_1,\ldots,x_n}$ and 
$\ket{y}=\ket{y_1,\ldots,y_n}$ are orthogonal if and only if 
$\ket{y_j}=\ket{x_j}^\perp$ for at least one index $j$. 
We use the abbreviation OPS to denote any set of pairwise 
orthogonal unit product vectors in $\cH$. 
The cardinality of an OPS cannot exceed $2^n$, the 
dimension of $\cH$. We say that an OPS is an OPB, 
{\em orthogonal product basis}, if its cardinality is $2^n$.
As an example, the $2^n$ product vectors 
$\ket{x_s}=\ket{s_1,\ldots,s_n}$, where 
$s:=(s_1,\ldots,s_n)$ runs through all binary 
$\{0,1\}$-sequences of length $n$, is an OPB. 
We refer to this OPB as the {\em standard OPB}. 
However, there are many other OPBs and describing or 
classifing them for any $n$ is a very hard problem. 
Let us also mention that a set of unit product vectors 
is called an {\em unextendible product basis} (UPB) if these 
vectors are orthogonal to each other and there is no product 
vector orthogonal to all of them \cite{bdm99,DiV03}. 
Originally it was required that UPB does not span the 
whole Hilbert space $\cH$, but for us it is convenient 
to drop that restriction. We shall say that a UPB is 
{\em proper} if it does not span $\cH$.

The above mentioned problem has been considered in 
\cite{bdm99,DiV03,bravyi,cj15,cd17,johnston13} and in our paper \cite{cd17JPA} where we studied 
the case $n=4$. For any $n$, we have reduced this 
classification problem to a purely combinatorial problem.
In order to extend this study further, we need to give
some basic definitions.

We start with an infinite countable alphabet $\bX$. We shall use 
letters (with indices if necessary) to denote the elements 
of this alphabet. The alphabet is equipped with a 
fixed-point-free involution $x\to x'$. Thus, for any 
$x\in\bX$ we have $x'\in\bX$, $x'\ne x$ and $(x')'=x$.
We shall refer to $x'$ as the {\em perpendicular} of $x$. 
We say that a subset of $\bX$ is {\em independent} if it does 
not contain any pair of the form $\{x,x'\}$. We shall also say that two vector varables $x$ and $y$ are independent if 
$x\ne y$ and $x\ne y'$. We shall refer to the letters $x\in\bX$ as {\em vector variables}. 

Let 
$x=[ ~ x_1 ~ x_2 ~ \cdots ~ x_n ~ ]$ and
$y=[ ~ y_1 ~ y_2 ~ \cdots ~ y_n ~ ]$ be two row vectors 
whose entries $x_j$ and $y_j$ are vector variables. 
We say that $x$ and $y$ are {\em orthogonal} to each other, 
$x\perp y$, if $y_j=x'_j$ for at least one index $j$.

Next, consider an $m\times n$ matrix $X=[x_{i,j}]$ 
with entries $x_{i,j}\in\bX$. 
We say that such a matrix is {\em orthogonal} if any two of 
its rows are orthogonal to each other. (Note that this is different from the traditional definition of orthogonal matrices in linear algebra).  
  
We denote by  $\cM(m,n)$ the set of all $m\times n$ matrices 
whose entries belong to $\bX$ and satisfy the following additional condition: if a vector variable $x$ occurs in some column of $X$ then neither $x$ nor $x'$ occurs in any other column of $X$. We denote by  $\cO(m,n)$ the subset of $\cM(m,n)$ consisting of all orthogonal matrices. We also set 
$\cO(n):=\cO(2^n,n)$ for the special case $m=2^n$. 
For two matrices $X_i\in\cM(m_i,n)$, $i=1,2$, we say they 
are {\em orthogonal to each other}, $X_1\perp X_2$, if each row 
of $X_1$ is orthogonal to each row of $X_2$.

If $X=[x_{i,j}]\in\cM(m,n)$ and $x\in\bX$ we define the 
{\em multiplicity}, $\mu(x,X)$, of $x$ in $X$ to be the number 
of pairs $(i,j)$ such that $x_{i,j}=x$. Thus if $x$ does not 
occur in $X$ then $\mu(x,X)=0$. When $X$ is known from the context we shall simplify this notation by writing just 
$\mu(x)$. It is easy to see that all maximal independent sets of 
vector variables, all of which occur in column $j$ of $X$, 
have the same cardinality. We denote this cardinality by 
$\nu_j(X)$. Finally, we set 
$\mu(X)=\max_{i,j} \mu(x_{i,j},X)$ and $\nu(X)=\sum \nu_j(X)$.

We say that a matrix $X\in\cM(m,n)$ is {\em balanced} if 
$\mu(x,X)=\mu(x',X)$ for all vector variables $x$. It is 
obvious that $X$ is not balanced if $m$ is odd. 
We have shown in \cite[]{cd17JPA} that all UOM in $\cO(n)$ are 
necessarily balanced.

Another important concept is the equivalence of matrices. 
We say that two matrices $X,Y\in\cM(m,n)$ are 
{\em equivalent} if $X$ can be transformed to $Y$ by 
permuting the rows, permuting the columns, and by renaming 
the vector variables. The renaming must respect the orthogonality, i.e., we require that if a vector variable $x$ is renamed to $y$ then $x'$ has to be renamed to $y'$. For $X\in\cM(m,n)$ we shall denote by $[X]$ its equivalence class. 
Note that if $X\in\cO(m,n)$ then $[X]\subseteq\cO(m,n)$.
Since $\bX$ is infinite, there are infinitely many matrices in $\cM(m,n)$. On the other hand, there are only finitely many equivalence classes in $\cM(m,n)$. 

Let us give a few very simple examples of orthogonal matrices. 
The set $\cO(1)$ has only one equivalence class, and $\cO(2)$ 
has two classes. Their representatives are
\bea \label{eq:primer1}
\bma 
a \\
a' \\
\ema, \quad 
\bma 
a & b \\
a & b' \\
a' & b \\
a' & b' \\
\ema, \quad 
\bma 
a & b \\
a & b' \\
a' & c \\
a' & c' \\
\ema.
\eea
(It is tacitly assumed that $a,b,c\in\bX$.)

\section{Evaluations and extensions of orthogonal matrices} 
\label{sec:extensions}

We say that a matrix $Y\in\cO(m,n)$ is an {\em extension} 
of $X\in\cO(p,n)$ if $X$ is a submatrix of $Y$. More precisely, 
we shall say that in this case $Y$ is a $k$-extension of $X$ 
where $k=m-p$. So the 0-extension (i.e., the trivial extension) of $X$ is $X$ itself. Note also that $X\in\cO(n)$ has no nontrivial extensions.
We say that a matrix $X\in\cO(m,n)$ is {\em extendible} 
if it has a 1-extension and otherwise we say that it is 
{\em unextendible}. We shall use the abbreviation UOM for 
``unextendible orthogonal matrix''. 
For example, every $X\in\cO(n)$ is a UOM. If $X\in\cO(m,n)$ is a UOM and $m<2^n$, then we shall say that $X$ is a {\em proper UOM}.

We say that two extensions $Y_1$ and $Y_2$ of $X\in\cO(m,n)$ 
are {\em equivalent extensions} of $X$ if the matrices $Y_1$ 
and $Y_2$ are equivalent. We shall count the number of extensions up to equivalence. 

To each orthogonal matrix we can associate a family of OPS. 
For this purpose we need the concept of evaluation of 
matrices.

\bd
\label{df:evaluation}
An {\em evaluation} of the matrix $X\in\cM(m,n)$ is a mapping $\alpha$ which assigns to each vector variable 
$x$, which occurs in $X$, a unit vector $\a(x)$ subject to the following two conditions:

(i) if $x$ occurs in column $j$ of $X$ (recall that such $j$ 
is unique) then $\a(x)\in\cH_j$;

(ii) if both $x$ and $x'$ occur in $X$ (necessarily in the 
same column) then $\a(x')=\a(x)^\perp$.

We say that the evaluation $\a$ of $X$ is {\em generic} if also 
the following condition is satisfied

(iii) if $x$ and $y$ are independent vector variables in the same column of $X$, then $\a(y)\ne\a(x)$ and $\a(y)\ne\a(x')$.
\qed
\ed

Note that the generic evaluations $\a$ are one-to-one, i.e., if $x\ne y$ then also $\a(x)\ne\a(y)$. Indeed, let $x$ and $y$ be two different vector variables which occur in $X$. If $x$ and $y$ occur in different columns of $X$ then $\a(x)\ne\a(y)$ because $\a(x)$ and $\a(y)$ belong to different Hilbert spaces $\cH_j$. Now assume that $x$ and $y$ occur in the same column of $X$. If they are independent, then (iii) implies that $\a(x)\ne\a(y)$. If $x$ and $y$ are dependent, then we have $y=x'$ and (ii) implies that $\a(y)=\a(x)^\perp\ne\a(x)$. 

Given an evaluation $\a$ of $X\in\cM(m,n)$, we obtain the 
$m\times n$ matrix $\a(X):=[\a(x_{i,j})]$ whose entries, 
$\a(x_{i,j})\in\cH_j$, are unit vectors. We shall refer to 
the matrix $\a(X)$ also as evaluation of $X$. Given this evaluation, we can form the product vectors
$\a(x_{i,1})\ox\cdots\ox\a(x_{i,n})$, $i=1,\ldots,m$. 
If the matrix $X$ is orthogonal, it is evident that these  
product vectors form an OPS of cardinality $m$. We refer to it 
as \textit{the OPS of $\a(X)$}. As the cardinality of an OPS cannot exceed 
$2^n$, we deduce that $\cO(m,n)$ is empty for $m>2^n$. 
If $X\in\cO(n)$ then the above OPS is in fact an OPB. 

To any $X\in\cO(m,n)$ we associate a family of OPS which arises from $X$ by applying all evaluations. 
We denote this family by $\cF_X$.
Further, we denote by $\cF^\#_X$ the subfamily of $\cF_X$ which arises from $X$ by applying only the generic evaluations. 
In the case $m=2^n$ these two families consist of OPBs.
For each $n$, up to equivalence there is exactly one matrix  
$X\in\cO(n)$ such that $\cF_X = \cF^\#_X$.
This is the matrix which in each column has just two vector variables, perpendicular to each other, and each with multiplicity $2^{n-1}$.

If $\a$ is any evaluation of an extendible orthogonal 
matrix $X$, then it is obvious that the OPS of $\a(X)$ 
is not a UPB. We consider next the case when $X$ is unextendible and $\a$ is generic.

\bl
\label{le:unext}
Let $\a$ be a generic evaluation of an UOM 
$X=[x_{i,j}]\in\cO(m,n)$. Then

(i) the OPS of $\a(X)$ is a UPB; 

(ii) each member of the family $\cF^\#_X$ is a UPB.
\el
\bpf
(i) Assume that the OPS of $\a(X)$ is not a UPB. Then there exists a product vector, say $\ket{b}=\ket{b_1}\ox\cdots\ox\ket{b_n}$, 
which is orthogonal to all product vectors 
$\ket{a_i}:=\a(x_{i,1})\ox\cdots\ox\a(x_{i,n})$, 
$i=1,2,\ldots,m$. 

We shall construct a new row $y=[~y_1~y_2~\cdots~y_n~]$ of 
vector variables. For each index $i$ there exists at least one index $j_i$ such that $\ket{b_{j_i}} \perp \a(x_{i,j_i})$. We assume that $j_i$ is chosen to be the smallest such index. 
Then we set $y_{j_i}:=x'_{i,j_i}$ for $i=1,\ldots,m$. 
It may happen that $j_i=j_k$ for some $k\ne i$. However in that case we have $\a(x_{i,j_i})=\a(x_{k,j_i})$, and so 
$x_{i,j_i}=x_{k,j_i}$ because $\a$ is generic. Thus the vector variable $y_{j_i}$ is well-defined for each index $j_i$. 
It may happen that the set of indices $J:=\{j_i:i=1,\ldots,m\}$ 
is a proper subset of $\{1,2,\ldots,n\}$. If $j\notin J$ then 
$y_j$ is taken to be a new variable independent from the 
variables which occur in $X$.

It is immediate from this construction that $y\perp X$. As $X$ 
is unextendible, we have a contradiction. We conclude that our assumption is false, i.e., the OPS of $\a(X)$ must be a UPB.

(ii) follows from (i) and the definition of $\cF^\#_X$.
\epf

The following lemma shows that each UPB arises as a generic 
evaluation of some UOM.
\bl
\label{le:upb-uom}
For each UPB, say $\cU$, there exists 
an UOM $X$ such that $\cU\in\cF^\#_X$. 
\el
\bpf
Let $\cU$ consist of $m$ product vectors 
$\ket{a_i}:=\ket{a_{i,1}}\ox\cdots\ox\ket{a_{i,n}}$, 
$i=1,2,\ldots,m$. 
We first construct the $m\times n$ matrix $A$ whose entries 
are the unit vectors $A_{i,j}:=\ket{a_{i,j}}$. Next, we choose an $m\times n$ matrix of independent vector variables $X=[x_{i,j}]$ on which we impose only the following relations:

(i) if $\ket{a_{i,j}}=\ket{a_{k,j}}$ then $x_{i,j}=x_{k,j}$; 

(ii) if $\ket{a_{i,j}}=\ket{a_{k,j}}^\perp$ then 
$x_{i,j}=x'_{k,j}$.

(We remind the reader that we consider two unit vectors as equal 
if they differ only in phase.)
Then $X\in\cO(m,n)$ and $\cU\in\cF^\#_X$. 
Lemma \ref{le:unext} implies that $X$ is unextendible.
\epf

In view of the two lemmas above many results about UPBs of 
multiqubit systems can be expressed in the language of UOMs
and vice versa.

For instance, our definition of equivalence in Sec. \ref{sec:pre} (restricted to the UOMs) is compatible with the 
definition of equivalence of the UPBs in multiqubit systems 
as defined in \cite[p. 4]{johnston14}. More precisely, let 
$X,Y\in\cO(m,n)$ be UOMs and let 
$\cU\in\cF_X^\#$ and $\cV\in\cF_Y^\#$. 
Then $[X]=[Y]$ (i.e., $X$ and $Y$ are equivalent) if and only if 
the UPBs $\cU$ and $\cV$ are equivalent.

A general necessary and sufficient condition for a set of orthogonal product states (in any finite-dimensional quantum system) to be unextendible has been obtained in 
\cite[Lemma 1]{DiV03}. In the case of qubit systems, that result has the following simple form.

\bl \label{le:DiV03}
A matrix $X\in\cO(m,n)$ is extendible if and only if there 
exist vector variables $y_1,\ldots,y_n$ such that, for each $j$, 
$y_j$ occurs in column $j$ of $X$ and each row of $X$ contains at least one of the $y_j$s.
\el

For each positive integer $n$ we denote by $\T_n$ the subset 
of $\{1,2,\ldots,2^n\}$ consisting of integers $m$ such that 
$\cO(m,n)$ contains at least one UOM. In view of 
Lemma \ref{le:unext} and the comments made above, $\T_n$ is just the set of all sizes of UPBs of $n$ qubits.
For instance, we have $\T_1=\{2\}$, $\T_2=\{4\}$ and 
$\T_3=\{4,8\}$, and note that $2^n\in\Theta_n$ for all $n$. 
In general, it is hard to determine whether $m\in\T_n$.

Let us give three examples of UOMs:
\begin{equation}
\label{eq:xy}
X=\left[\begin{array} {ccc}
a & c & e  \\
a'& d'& f  \\
b & c'& f' \\
b'& d & e' \end{array} \right], \quad
Y=\left[\begin{array} {cccc}
x  & y  & z  & w  \\
x' & b  & d  & e  \\
a  & y' & d' & f  \\
a' & c  & z' & e' \\
a  & b' & d  & w' \\
x' & c' & d' & f' \end{array} \right], \quad
Z=\left[\begin{array} {cccccccc}
a  & e  & i  & m  & q  & v  & \a  & \z   \\
b  & f  & j  & m' & r  & w  & \b  & \eta \\
c  & g  & k  & n  & s  & v' & \b' & \t   \\
a' & f' & l  & n' & t  & x  & \g  & \i   \\
c' & h  & j' & o  & t' & y  & \a' & \k   \\
b' & h  & i' & p  & s' & y' & \d  & \i'  \\
b' & h' & l' & n' & u  & z  & \e  & \z'  \\
c' & h' & i' & p  & r' & z' & \d  & \i'  \\
a' & f' & l' & p' & t  & x  & \e' & \t'  \\
d  & g' & l  & p' & q' & w' & \g' & \k'  \\
d' & e' & k' & o' & u' & x' & \d' & \eta'
\end{array} \right].
\end{equation}
These matrices are obtained from the known three-qubit, four-qubit and eight-qubit UPBs (see \cite{bravyi,fs09,johnston13}). The UOMs $X$ and $Y$ are unique up to equivalence \cite{johnston14}, but it is not known whether $Z$ is unique. Further, $X$ is balanced while $Y$ and $Z$ are not.

In the following lemma we collect some basic properties of UOMs.

\bl
\label{le:upb1}
Let $X=[x_{i,j}]\in\cO(m,n)$ be a UOM. Then

(i) $x'_{i,j}$ occurs in $X$ for all $i,j$;

(ii) if the vector variables $y_1,y_2,\ldots,y_s$ occur in $s$ 
different columns of $X$, then $X$ has at least $n-s+1$ 
rows containing no $y_j$; 

(For $s=0$ we obtain that $m>n$, and for $s=1$ we obtain that 
$\mu(X)\le m-n$.)

(iii) for each $i$, $\sum_{j=1}^n \mu(x'_{i,j})\ge m-1$ and the equality holds if and only if $x_{k,j}=x'_{i,j}$ implies that 
$x_{k,s}\ne x'_{i,s}$ whenever $s\ne j$;

(iv) there is an index $r$ such that $x_{r1}=x'_{11}$ and 
$x_{rs}\ne x'_{1s}$ for $s>1$;

(v) if $Y\in\cO(d,k)$, $k<n$, is a submatrix of $X$ then 
$d\le m-n+k$ and for $k \ge n/2$ we have $d<m-n+k$; 

(vi) if $p_j=\sum \mu(x)\mu(x')$, where the summation is over all pairs $\{x,x'\}$ in column $j$ of $X$, then $\sum p_j \ge m(m-1)/2$.
\el
\bpf
(i) Assume that say $\mu(x'_{1,1})=0$. Then 
$[~x'_{1,1}~x_{1,2}~\cdots~x_{1,n}~] \perp X$, and we have a contradiction.

(ii) Denote by $r$ the number of rows of $X$ containing no 
$y_j$. Suppose that $r\le n-s$. By permuting the rows and columns of $X$, we may assume that $y_j$ occurs in column 
$j$ for each $j$ and that the first $r$ rows contain no $y_j$.
Then the row 
$[~y'_1~\cdots~y'_s~x'_{1,1+s}~\cdots~x'_{r,r+s}~x_{1,r+s+1}
~\cdots~x_{1,n}~]$ is orthogonal to each row of $X$. As $X$ is a UOM, we have a contradiction. We conclude that $r>n-s$.

(iii) The inequality follows from the fact that the row $i$ of 
$X$ is orthogonal to each of the other $m-1$ rows. 

(iv) Let us write the first row of $X$ as $v:=[~x_{11}~u~]$, 
where $u=[~x_{12}~\cdots~x_{1n}~]$, and set $w:=[~x'_{11}~u~]$. 
If $w$ is a row of $X$, we can take $r$ to be the index of that 
row. In that case the assertion obviously holds.
We may now assume that $w$ is not a row of $X$. Since $X$ has 
no 1-extensions, there is a row of $X$, say row $r$, which 
is not orthogonal to $w$. Therefore $x_{r1}\ne x_{11}$ and $u$ is not orthogonal to $[~x_{r2}~\cdots~x_{rn}~]$. Hence, 
$r>1$ and the orthogonality of rows 1 and $r$ of $X$ implies 
that $x_{r1}=x'_{11}$. This completes the proof.

(v) We may assume that $X=\bma Y & P \\ Q & R \\ \ema$. 
Let $y$ and $p$ be the first rows of $Y$ and $P$, respectively. 
If $d>m-n+k$ then $m-d<n-k$ and we can choose a row $z$ of 
length $n-k$ which is orthogonal to $p$ and $R$.
Then $[~y~z~] \perp X$, which gives a contradiction. We conclude that $d\le m-n+k$.

Suppose now that $k \ge n/2$. Assume that $d=m-n+k$. 
If a row of $Y$ is orthogonal to a row of $Q$, say the first 
rows $y$ and $q$ of these matrices, then there exists a row
$z$ of length $n-k$ which is orthogonal to the first row $p$ of $P$ and all the rows of $R$ except the first. Then 
$[~q~z~] \perp X$ and we have a contradiction.
We conclude that no row of $Y$ is orthogonal to a row of $Q$. 
As $X$ is orthogonal, we must have $Q\perp R$. Since $Q$ has 
$n-k$ rows and $n-k\le k$, there exists a row $s\perp Q$. 
Hence, if $r$ is a row of $R$ then $[~s~r~] \perp X$ and we have a contradiction. We conclude that $d<m-n+k$.

(vi) This follows from the fact that for each pair $(i,k)$, 
$i<k$, the rows $i$ and $k$ of $X$ are orthogonal to each other.
See also \cite[Appendix A]{johnston14}.
\end{proof}

We remark that Lemma \ref{le:upb1} (i) is equivalent to 
\cite[Lemma 2]{johnston13}. 
The inequality $\m(X)\le m-n$ mentioned in part (ii) of Lemma \ref{le:upb1} has been proved in \cite[Appendix A, p. 16]{johnston14}.


\bcr \label{cr:mu=1}
If $X\in\cO(n+1,n)$, $n$ odd, then $X$ is a UOM if and only if 
$\mu(X)=1$. 
\ecr
\bpf
The condition is necessary by Lemma \ref{le:upb1} (iii). 

Suppose that $\mu(X)=1$. Assume that there exists a row 
$y=[~y_1~y_2~\cdots~y_n~]\perp X$. Then we must have 
$\sum\mu(y'_j,X)\ge n+1$. As each $\mu(y'_j,X)\le1$, we have a contradiction. Hence, there is no such $y$ which means that 
$X$ is a UOM. 
\epf

Let $A\in\cM(r,s)$ and $B\in\cM(m,n-s)$, $n>s$, and assume 
that any vector variables $x$ and $y$ of $A$ and $B$, 
respectively, are independent. Further, let $B$ be 
partitioned into blocks $B_k\in\cM(m_k,n-s)$, $k=1,\ldots,r$, 
$$
B=\left[ \begin{array}{c} B_1 \\ \vdots \\ B_r \end{array}
\right].
$$
Given such data, we denote by
$$
A\models(B_1,B_2,\ldots,B_r)
$$ 
the matrix $[~\tilde{A}~B~]\in\cM(m,n)$, where $\tilde{A}$ 
is the $m\times s$ matrix obtained from $A$ by replacing, 
for each $k$, the row $k$ of $A$ by $m_k$ copies of that row.
Note that if a vector variable $x$ occurs in one of the 
blocks $B_k$ then $x$ or $x'$ may occur in another block 
but necessarily in the same column.

It is straightforward to verify that if $A$ and the $B_k$ 
are orthogonal matrices, then the matrix 
$A\models(B_1,B_2,\ldots,B_r)$ is also orthogonal. 

Let us give an  example. We take $r=4$, $s=3$, $n=5$ and set 
$$
A=\left[\begin{array} {ccc}
a  & c  & e  \\
a' & d' & f  \\
b  & c' & f' \\
b' & d  & e' \end{array} \right], \quad
B_k=
\bma
g  & h  \\
g  & h' \\
g' & h  \\
g' & h' \\
\ema,~ k=1,2,3, \quad
B_4=
\bma
x  & y  \\
x  & y' \\
x' & z  \\
x' & z' \\
\ema.
$$

Then we have 
\bea
A\models(B_1,B_2,B_3,B_4)=
\left[\begin{array} {ccccc}
a & c & e  & g  & h  \\
a & c & e  & g  & h' \\
a & c & e  & g' & h  \\
a & c & e  & g' & h' \\
a'& d'& f  & g  & h  \\
a'& d'& f  & g  & h' \\
a'& d'& f  & g' & h  \\
a'& d'& f  & g' & h' \\
b & c'& f' & g  & h  \\
b & c'& f' & g  & h' \\
b & c'& f' & g' & h  \\
b & c'& f' & g' & h' \\
b'& d & e' & x  & y  \\
b'& d & e' & x  & y' \\
b'& d & e' & x' & z  \\
b'& d & e' & x' & z' \\  \end{array} \right].
\eea
In this case $A$ and the $B_k$ are orthogonal, and so the matrix 
$A\models(B_1,B_2,B_3,B_4)$ is also orthogonal.

\bpp
\label{pp:konstr}
Let $X:=A\models(B_1,B_2,\ldots,B_r)$ where $A\in\cO(r,s)$, 
$B_k\in\cO(m_k,n-s)$ for $k=1,2,\ldots,r$. 
Then $X$ is a UOM if and only if $A$ and all the $B_k$ are UOMs.
\epp
\bpf
Assume that $X$ is a UOM. If $A$ is not a UOM, choose a row 
$u\in\cM(1,s)$ such that $u \perp A$. 
If $v$ is the first row of $B_1$ then row $[~u~v~]$ is orthogonal to $X$. Thus we have a contradiction. 
Similarly, we get a contradiction if at least one $B_k$ 
is not a UOM.

Now assume that $A$ and all the $B_k$ are UOMs. Let 
$[~u~v~]\in\cM(1,n)$ with $u\in\cM(1,s)$. Since $A$ is a UOM, 
$u$ is not orthogonal to some row of $A$, say the first row. 
Since $B_1$ is a UOM, $v$ is not orthogonal to some row of 
$B_1$, say the first row. Thus $[~u~v~]$ is not orthogonal 
to the first row of $X$. Hence, $X$ is not extendible, i.e., 
$X$ is a UOM.
\epf

\bcr
\label{cr:rts}
If $r\in\T_s$ and $m_1,m_2,\ldots,m_r\in\T_{n-s}$ then 
$\sum m_i \in \T_n$.
\ecr

For example, as $2\in\T_1$ and $4,8\in\T_3$ we obtain that
$8,12\in\T_4$.

Two extreme cases of this corollary are used often. The first 
case is $s=1$, which implies that $r=2$. It has been used 
extensively by Johnston \cite[Proposition 1]{johnston14}. 
The other extreme case is $s=n-1$, which implies that all 
$m_k=2$.

We say that a matrix $X\in\cM(m,n)$ is {\em decomposable} 
if it is equivalent to a matrix $A\models(B_1,B_2,\ldots,B_r)$, 
$r\ge 1$. 

We exhibit below two UOMs in $\cO(8,4)$, the first one is decomposable while the second one is not. 

\begin{equation} \label{eq:primeri4q} 
\left[ \begin{array}{cccc}
x  & a  & c  & e  \\
x  & a' & d  & f  \\
x  & b  & c' & f' \\
x  & b' & d' & e' \\
x' & g  & i  & k  \\
x' & g' & j  & l  \\
x' & h  & i' & l' \\
x' & h' & j' & k'
\end{array} \right], \quad
\left[ \begin{array}{cccc}
a  & c  & e  & g  \\
a  & d  & e' & h  \\
a  & d' & f  & g' \\
a' & c' & f  & h  \\
a' & c  & e  & g  \\
b  & d' & f' & g' \\
b  & d  & e' & h' \\
b' & c' & f' & h'
\end{array} \right]. 
\end{equation}

We give below an example of an orthogonal $17\times5$ matrix 
which is orthogonal to 16 rows but has no extensions in $\cO(5)$.

\begin{example} {\rm
Let $Y\in\cO(6,4)$ be the UOM given in \eqref{eq:xy}, and $Z$ any UOM in $\cO(12,4)$ having no vector variables in common with $Y$. Then the matrix $X:=\bma s\\s'\ema\models(Y,Z)\in\cO(18,5)$ is a UOM ($s$ is a new vector variable). 
Let $P$ be the submatrix of $X$ obtained by removing the first row. So $P=\bma s\\s'\ema\models(Q,Z)\in\cO(17,5)$ where 
$$
Q=\left[\begin{array}{cccc}
x' & b  & d  & e  \\
a  & y' & d' & f  \\
a' & c  & z' & e' \\
a  & b' & d  & w' \\
x' & c' & d' & f' \end{array} \right].
$$
There are exactly 16 rows orthogonal to $Q$, say 
$q_1,\ldots,q_{16}$. Let us list these rows:

\begin{eqnarray*} 
&& [~a'~b'~z~f~],~[~a'~b'~d~e~],~[~a'~c'~d'~f~],~[~a'~c'~d~e'~],\\
&& [~a'~c~d'~e~],~[~a'~c~z~e'~],~[~x~y~z~w~],~[~x~y~d'~e~],\\
&& [~x~c'~d~w~],~[~x~c'~d'~f'~],~[~x~b~d~e~],~[~x~b~z~f'~],\\
&& [~a~y~d'~f~],~[~a~c~d'~f'~],~[~a~b'~d~w~],~[~a~b~d~e'~].
\end{eqnarray*}

It follows that among the $q_j$s there are 
at most 11 mutually orthogonal rows. There are also exactly 16 rows orthogonal to $P$, namely the rows $[~s~q_j~]$. Hence, at 
most 11 of these extended rows may be mutually orthogonal. 
Since $17+11=28<32$, $P$ has no extensions in $\cO(5)$. }
\end{example}

We shall need later the following simple lemma.

\bl \label{le:extpart}
Let $X\in\cO(m,n)$ be partitioned as $X=[~X_1~X_2~]$, where 
$X_1\in\cO(m,s)$. Suppose that $X_1$ has a $p$-extension 
$Y_1$. Then $X$ has a $k$-extension $Y$ with 
$k=2^{n-s}(m+p)-m $.
\el
\bpf
We may assume that $Y_1=\left[ \begin{array}{c}
X_1 \\ U \end{array} \right]$.
For each $i\in\{1,2,\ldots,m+p\}$ choose a matrix 
$Z_i\in\cO(n-s)$.
If $i\le m$ we choose $Z_i$ so that its first row is equal 
to the row $i$ of $X_2$. Let $Z$ be the $2^{n-s}(m+p)\times(n-s)$ matrix obtained by stacking the matrices 
$Z_1,Z_2,\ldots,Z_{m+p}$ one on top of the other. 
Let $\tilde{Y_1}$ be the $2^{n-s}(m+p)\times s$ matrix obtained 
from $Y_1$ by replacing each row by $2^{n-s}$ copies of it. 

We claim that the $2^{n-s}(m+p)\times n$ matrix 
$Y=[~\tilde{Y_1}~Z~]$ is orthogonal. To prove this claim, 
let $u:=[~a~b~]$ and $v:=[~c~d~]$ be two rows of $Y$, where $a$ and $c$ have length $s$. Since no two rows of $Y$ are equal, 
we have $a\ne c$ or $b\ne d$. 
If $a\ne c$ then $a$ and $c$ are orthogonal since they are 
two rows of $Y_1$. If $a=c$ then $b\ne d$ and $b$ and $d$ are orthogonal since they are two rows of some $Z_i$. 
We conclude that $u$ and $v$ are orthogonal and our claim is proved.

Finally, it is easy to verify that $X$ is a submatrix of $Y$.
\epf

Note that if $p=2^s-m$ then $k=2^n-m$, i.e., if $Y_1\in\cO(s)$ 
then $Y\in\cO(n)$. 

Let us give an example with $m=n=3$, $s=2$, $p=1$, $k=5$:
$$
X=\left[\begin{array}{ccc}
a &b &c \\ a' &d &e \\ a &b' &e'
\end{array}\right], \quad
X_1=\left[\begin{array}{cc}
a &b \\ a' &d \\ a &b' 
\end{array}\right], \quad
X_2=\left[\begin{array}{c}
c \\ e \\ e' 
\end{array}\right].
$$
Then we have
$$
Y_1=\left[\begin{array}{cc}
a &b \\ a' &d \\ a &b' \\ a' &d'
\end{array}\right], \quad
Z_1=\left[\begin{array}{cc}c\\c'\end{array}\right],\quad
Z_2=\left[\begin{array}{cc}e\\e'\end{array}\right],\quad
Z_3=\left[\begin{array}{cc}e'\\e\end{array}\right],\quad
Z_4=\left[\begin{array}{cc}x\\x'\end{array}\right],
$$
where $x$ could be a new vector variable or one of 
$c,c',e,e'$. Finally,
$$
Y_1=\left[\begin{array}{cc}
a &b \\ a &b \\ a' &d \\ a' &d \\ 
a &b' \\ a &b' \\ a &d' \\ a &d'
\end{array}\right], \quad
Y=\left[\begin{array}{ccc}
a &b &c \\ a &b &c' \\ a' &d &e \\ a' &d &e' \\ 
a &b' &e' \\ a &b' &e \\ a &d' &x \\ a &d' &x'
\end{array}\right].
$$

\section{Old and new facts about multiqubit UPBs}
\label{sec:facts}

The problem of computing $\Theta_n$ has been considered by several authors \cite{AL01,kf06,bravyi}. In the following theorem we list the main facts presently known about $\T_n$. For these we refer to the papers of Di Vincenzo et al. 
\cite{DiV03}, Feng \cite{kf06} and Johnston \cite[Theorems 1]{johnston13} and \cite[Propositions 2,3 and Theorems 4,5]{johnston14}. 
The fact that $2^n-5\notin\T_n$ has been proved recently in 
\cite[Theorem 1]{cd17}. 

We set $\theta_n=\min\T_n$, the smallest integer of $\T_n$.

\bt
\label{thm:upb}
(i) $\theta_n=n+1$ if $n$ is odd;
$\theta_n=n+2$ if $n\equiv2 \pmod{4}$;
$\theta_n=n+4$ if $n\equiv0 \pmod{4}$ and $n>8$;
$\theta_4=6$ and $\theta_8=11$.

(ii) $n+2\notin\T_n$ if $n$ is odd.

(iii) $2^n\in\T_n$ for all $n$;  $2^n-4\in\T_n$ for $n\ge3$; and 
$2^n-k\notin\T_n$ for all $n$ and $k=1,2,3,5$.

(iv) $m\in\T_n$ if $n\ge7$ and 
$\sum_{k=4}^{n-1} \theta_k \le m\le 2^n-6$.

(v) $m\in\T_n$ if $n<m\le2^n$ and $m\equiv0 \pmod{4}$,
except for the case where $n\equiv1 \pmod{4}$ and $m=2n+2$ which in general remains undecided.
\et

The sum $\sum_{k=4}^{n-1} \theta_k$ is very closely approximated 
by the quadratic polynomial $n(n+3)/2-15$. It is shown in \cite{johnston14} that the differences 
$$
\left( n(n+3)/2-15 \right)-\sum_{k=4}^{n-1}\theta_k,\quad n\ge7,
$$
belong to $\{0,1,2\}$. One can easily verify that for $n\ge9$ this sequence is periodic with period $0,1,1,2$.

In Table \ref{tab:sizes} we give the UOM of sizes $11\times 5$, $10\times 6$, $11\times 6$, $13\times 6$ and $m\times 7$ for $m=13,14,15,19$. As far as we know, no such UOM have been discovered so far. Another example of $19\times7$ UOM, not 
equivalent to the one in this table, can now be constructed 
by using Corollary \ref{cr:rts}. Indeed, we know that 
$8,11\in\T_6$ and so $19\in\T_7$.
Let us explain the notation used in Table \ref{tab:sizes}.
The same lower and upper case letters, say ``a'' and 
``A'', are perpendiculars of each other, i.e., $A=a'$. 
We write the matrix by listing its rows in order. To save 
space we omit blanks between the letters in the same row. 
All entries in column, say $j$, should be adorned with
the subscript $j$, so that no variable occurs in two different 
columns.

From the above theorem we deduce the following simple corollary.
\bcr \label{cr:uomexist}
We have $\T_n\subseteq\T'_n$ where

(a) $\T'_n=\{n+1,n+3,n+4,\ldots,2^n-6,2^n-4,2^n\}$ for $n$ odd;

(b) $\T'_n=\{\theta_n,\theta_n+1,\ldots,2^n-6,2^n-4,2^n\}$ 
for $n$ even.
\ecr

Since it is known that $\Theta_1=\{2\}$, $\Theta_2=\{4\}$, $\Theta_3=\{4,8\}$ and $\T_4=\{6-10,12,16\}$, 
see \cite[Table 3]{johnston14}, we have $\T_n=\T'_n$ for 
$n\le 4$. Since we have constructed UOMs of new sizes: 
$11\times 5$, $m\times 6$ for $m\in\{10,11,13\}$, and 
$m\times 8$ for $m\in\{12-15,17-19\}$ 
(see Sec. \ref{app:B} Table \ref{tab:sizes}), 
we have also $\T_5=\T'_5$, $\T_6=\T'_6$ and $\T_8=\T'_8$. 
(The abbreviation $i-j$ stands for the sequence $i,i+1,\ldots,j$ 
of all integers in the range from $i$ to $j$.)
We were not able to decide whether 10 or 11 belongs to $\T_7$.

For other $n\le13$, we list in Table \ref{tab:decided} the integers known to belong to $\T_n$. 

\begin{table}[h] 
$$
\begin{array}{cclc}
n & ~ & \qquad\qquad\qquad m & \\
\hline \\
7 && 8,12-122,124,128  \\
9 && 10,12,16,22-506,508,512 \\
10 && 12,16,20-22,24,26,28,32-1018,1020,1024 \\
11 && 12,16,20,24,28,32,34,36,38,40,42,44-2042,2044,2048 \\
12 && 16,20,24,28,32,36,40,42,44,46,48,50-4090,4092,4096 \\
13 && 14,16,20,24,32,36,40,44,48,52,56,58,60-8186,8188,8192 \\
\end{array}
$$
\caption{Known cases of $m\in\T_n$, $n=7,9-13$.}
\label{tab:decided}
\end{table}

%

Very little is known about the number of UOM equivalence classes 
in $\cO(m,n)$, $m\in\T_n$. 
In Table \ref{tab:equiv-classes} we record the known facts 
for $m\le12$. (The blanks occur outside the range 
$\theta_n \le m\le 2^n$.)
The main reference is Johnston's paper 
\cite{johnston14} where the electronic links to his 
computational results are provided. 

\begin{table}[h] 
$$
\begin{array}{r|rrrrrrrrrrr}
m \backslash n~ &1&2&3&4&5&6&7&8&9&10&11\\
\hline \\
 1&&&&&&&&&&&\\
 2&1&&&&&&&&&&\\
 3&&&&&&&&&&&\\
 4&&2&1&&&&&&&&\\
 5&& &0&&&&&&&&\\
 6&& &0&    1& 1&&&&&&\\
 7&& &0&    1& 0&&&&&&\\
 8&& &17& 144&32&9&6&&&&\\
 9&& &  &  11& e&e&0&&&&\\
10&& &  &  80& e&e&?& &e&&\\
11&& &  &   0& e&e&?&e&0&&\\
12&& &  &1209& e&e&e&e&e&e&e\\
\end{array}
$$
\caption{Number of UOM equivalence classes. The letter $e$ means that 
the UOMs exist in $\cO(m,n)$. The two question marks 
indicate that the existence of UOMs is still undecided.}
\label{tab:equiv-classes}
\end{table}

For instance, there is only 1 UOM-equivalence class in 
$\cO(2,1)$, only 2 UOM-equivalence classes in $\cO(4,2)$, 
and only 1 UOM-equivalence class in $\cO(4,3)$. 
The two non-equivalent UOMs in $\cO(2)$ are exhibited in 
\eqref{eq:primer1}. 

Most of these results have been obtained by computer searches. 
Some are easy to verify. For instance let us verify the claim for $(m,n)=(4,3)$. 
Let $X=[x_{i,j}]\in\cO(4,3)$. By Lemma \ref{le:upb1} (ii) 
we have $\mu(X)=1$, i.e. $\mu(x_{i,j})=1$ for all $i,j$.
By using Lemma \ref{le:upb1} (i), we deduce that each column 
of $X$ must contain exactly two independent vector variables 
and their perpendiculars. By permuting the rows of 
$X$, we can assume that $x_{2,1}=x'_{1,1}$, $x_{4,1}=x'_{3,1}$ 
and $x_{1,2}=x'_{3,2}$. It follows that $x_{2,2}=x'_{4,2}$. 
Since the first and fourth rows are orthogonal, we have 
$x_{1,4}=x'_{4,4}$. Hence, we must have $x_{2,4}=x'_{3,4}$ 
and so $X$ is equivalent to the first matrix in \eqref{eq:xy}.

Up to column permutations, a matrix $X=[x_{ij}]\in\cO(m,n)$ can be replaced by the set $\{X_1,X_2,\ldots,X_n\}$, where the temporary symbol $X_r$ represents the column $r$ of $X$. 
Let $k$ be the number of independent vector variables in column $r$, i.e., $k=\nu_r=\nu_r(X)$. We choose a maximal independent set $\{a_1,a_2,\ldots,a_k\}$ of vector variables in column $r$. 
For each $s\in\{1,2,\ldots,k\}$ we define the index sets  
$\Omega'_{s,r}=\{i:x_{ir}=a_s\}$ and 
$\Omega''_{s,r}=\{i:x_{ir}=a'_s\}$, and we set 
$\Omega_{s,r}=\{\Omega'_{s,r},\Omega''_{s,r}\}$.
Finally we replace the temporary symbol $X_r$ with the set
$\Omega_r:=\{\Omega_{1,r},\ldots,\Omega_{k,r}\}$. 
We shall refer to the set $\Omega:=\{\Omega_1,\ldots,\Omega_n\}$ as the {\em symbol} of $X$. 

For instance, for the first matrix in \eqref{eq:primeri4q} we have
\bea
\Omega=\{\{\Omega_{1,1},\Omega_{2,1}\},\{\Omega_{1,2},\Omega_{2,2}\},
\{\Omega_{1,3},\Omega_{2,3}\},\{\Omega_{1,4},\Omega_{2,4},\Omega_{3,4}\}\},
\eea
where
\begin{eqnarray*}
&& \Omega_{1,1}=\{\{1,2,3\},\{4,5\}\},
~~~~~~
 \Omega_{2,1}=\{\{6\},\{7\}\},\\
&& \Omega_{1,2}=\{\{1,4\},\{2,6\}\},
~~~~~~~~~
 \Omega_{2,2}=\{\{3,5\},\{7\}\},\\
&& \Omega_{1,3}=\{\{1,4\},\{7\}\},
~~~~~~~~~~~~
 \Omega_{2,3}=\{\{2,5\},\{3,6\}\},\\
&& \Omega_{1,4}=\{\{1,6\},\{3\}\},
~~~~~~~~~~~~
 \Omega_{2,4}=\{\{2\},\{7\}\},
~~~~~~~~~~~
 \Omega_{3,4}=\{\{4\},\{5\}\}.
\end{eqnarray*}

Since we assumed only that $X\in\cO(m,n)$, it may happen that 
some of the index sets $\Omega''_{s,r}$ are empty. However, 
this is not the case if $X$ is an UOM.
Note that the symbol $\Omega$ is uniquely determined by $X$. 
It does not depend on the choice of the maximal sets 
$\{a_1,a_2,\ldots,a_k\}$ of independent vector variables in a 
column. It also does not depend on the ordering of these variables. 
Moreover, it is affected by neither column permutations nor 
renaming of the vector variables. 
The row permutations of $X$ do change the symbol $\Omega$.
Their effect consists in permuting the integers $1,2,\ldots,m$. 
More precisely, if $\pi$ is a permutation of these integers, then $\pi\Omega$ is obtained from $\Omega$ by simultaneously 
replacing each integer $i$ by $\pi(i)$ (all $n$ occurencies 
of $i$ in $\Omega$). 

We can recover $X$ up to equivalence from its symbol $\Omega$. To do this, for $r\in\{1,2,\ldots,n\}$ let $k$ be the cardinality of the set 
$\Omega_r:=\{\Omega_{1,r},\ldots,\Omega_{k,r}\}$. 
We select a set of independent vector variables $\{z_1,z_2,\ldots,z_k\}$ and define the column $r$ of the matrix $Y=[y_{ij}]\in\cM(m,n)$ by setting 
$y_{ir}=z_s$ if $i\in\Omega'_{s,r}$ and 
$y_{ir}=z'_s$ if $i\in\Omega''_{s,r}$. 
Then the matrices $X$ and $Y$ are equivalent. 
(The two choices for the ordered pair 
$(\Omega'_{s,r},\Omega''_{s,r})$ 
give equivalent matrices $Y$.)

Thus we obtain the following simple test for 
equivalence of two UOMs.

\bl \label{le:eq-test}
Two UOMs $X$ and $Y$ in $\cO(m,n)$ with symbols $\Omega_X$ and $\Omega_Y$, respectively, are equivalent if and only if 
$\Omega_Y=\pi\Omega_X$ for some permutation $\pi$ of 
$\{1,2,\ldots,m\}$.
\el

(Note that this test is not efficient for large values of $n$.)

\section{New construction of UOM}
\label{sec:newconstr}

Let $X\in\cM(2m,n)$ be partitioned into two blocks, $X_1$ and 
$X_2$, of the same size, and assume that $X_1,X_2\in\cO(m,n)$.
From $X$ we can construct a matrix $Z\in\cO(2m,m+n)$. 

As the first step of our construction, we choose a matrix 
$Y_1:=[y_{ij}]$ of order $m$ where $y_{ij}$ are new vector variables which are mutually independent and also independent 
from the varables that occur in $X$. 

The second step is to construct another matrix of order $m$. 
This matrix, $Y_2$,  is obtained from $Y_1$ by first replacing each entry by its perpendicular and then, 
for each $j\in\{1,2,\ldots,m\}$, rotating the entries in the 
$j$th column $j-1$ steps downwards. 

For example, if $m=4$ we have
\bea \label{eq:Y_1,Y_2}
Y_1=
\bma
y_{11} & y_{12} & y_{13} & y_{14} \\
y_{21} & y_{22} & y_{23} & y_{24} \\
y_{31} & y_{32} & y_{33} & y_{34} \\
y_{41} & y_{42} & y_{43} & y_{44} 
\ema, \quad
Y_2=
\bma
y'_{11} & y'_{42} & y'_{33} & y'_{24} \\
y'_{21} & y'_{12} & y'_{43} & y'_{34} \\
y'_{31} & y'_{22} & y'_{13} & y'_{44} \\
y'_{41} & y'_{32} & y'_{23} & y'_{14} 
\ema.
\eea

From the definition of $Y_2$ it follows that for each $i$ no two 
of the vector variables $y'_{i,1}, y'_{i,2},\ldots,y'_{i,m}$ 
occur in the same row of $Y_2$. Thus each row of $Y_1$ is perpendicular to each row of $Y_2$. This property of the matrices 
$Y_1$ and $Y_2$ is essential for our construction.

Finally we define our matrix $Z$ by setting 
\begin{equation}
\label{eq:xy1}
Z=\bma
X_1 & Y_1 \\
X_2 & Y_2
\ema.	
\end{equation}

The fact that $Z\in\cO(2m,m+n)$ follows immediately from the above mentioned property and the hypothesis that 
$X_1,X_2\in\cO(m,n)$. 

For the sake of brevity, we shall write $X\LRa Z$ or 
$(X_1,X_2)\LRa Z$ to indicate that $Z$ is obtained from $X$ by 
the above construction, and by choosing the matrix $Y_1$ 
appropriately.

Let us assume now that $n$ is odd. In that case $\theta_n=n+1$. 
The $n$-qubit UPBs of cardinality $n+1$, known as GenShift UPBs, have been constructed in \cite[p. 395]{DiV03}. They give UOMs in $\cO(n+1,n)$, which will be called {\em GenShift UOMs}.

Let us recall that construction. Let $n=2p+1$ and denote by $Z$ 
the $n+1$ by $n$ matrix obtained from the cyclic matrix of order $n$, with first row $[~0~1'~2'~\cdots~p'~p~\cdots~2~1~]$, 
by appending at the bottom the row having all entries equal to 
$0'$. 
Next for each $j\in\{1,2,\ldots,n\}$ choose $p+1$ independent vector variables $y_{i,j}$, $i=0,1,\ldots,p$. Finally, construct $X$ from $Z$ by replacing, for each $j$, the entries in column $j$ by vector variables as follows: 
$k\to y_{k,j}$ and $k'\to y'_{k,j}$, $k\in\{0,1,\ldots,p\}$. 

Obviously, the last row of $X$ is orthogonal to all other rows.
Let us verify that the rows $i$ and $j$, $1\le i<j\le n$, of $X$ are orthogonal. 
If $j-i$ is even, this follows from $Z_{i,i+k}=k'$ and 
$Z_{j,j-k}=k$, where $k=(j-i)/2$. 
If $j-i$ is odd, it follows from $Z_{i,i-k}=k$ and 
$Z_{j,j+k}=k'$, where $k=p-(j-i-1)/2$. 
(The second subscript should be reduced modulo $n$ to
be in the range $1,2,\ldots,n$.)
As no column of $Z$ has two equal entries, 
Corollary \ref{cr:mu=1} implies that $X$ is an UOM. 

Let us give an example. We take $n=5$, and choose 15 independent vector variables $y_{i,j}$, $i=0,1,2$; $j=1,\ldots,5$. Then the matrix $Z$ and the corresponding matrix $X$ are

\bea
\label{le:zx}
Z=\bma
0 & 1'& 2'& 2 & 1 \\
1 & 0 & 1'& 2'& 2 \\
2 & 1 & 0 & 1'& 2' \\
2'& 2 & 1 & 0 & 1' \\
1'& 2'& 2 & 1 & 0 \\
0'& 0'& 0'& 0'& 0'
\ema,
\quad
X=\bma
y_{01} & y_{12}'& y_{23}'& y_{24} & y_{15} \\
y_{11} & y_{02} & y_{13}'& y_{24}'& y_{25} \\
y_{21} & y_{12} & y_{03} & y_{14}'& y_{25}'\\
y_{21}'& y_{22} & y_{13} & y_{04} & y_{15}'\\
y_{11}'& y_{22}'& y_{23} & y_{14} & y_{05} \\
y_{01}'& y_{02}'& y_{03}'& y_{04}'& y_{05}'
\ema.
\eea

We present now our alternative construction of UOMs in 
$\cO(n+1,n)$ when $n=2p+1$ and $p$ is odd. Since $\theta_p=p+1$, there exist UOMs in $\cO(p+1,p)$ e.g. the above GenShift UOM.

\bpp \label{pp:novuom}
Let $X_1,X_2\in\cO(p+1,p)$ be UOMs having no vector variable in common and let 
$$
(X_1,X_2)\LRa Z=\bma
X_1 & Y_1 \\
X_2 & Y_2
\ema	
\in\cO(2p+2,2p+1).
$$
Then the matrix $Z$ is a UOM.
\epp
\bpf
From the construction of $Z$ we know that $Z$ is an orthogonal 
matrix. It is easy to see that $\mu(Z)=1$. Hence, $Z$ is a UOM by Corollary \ref{cr:mu=1}. 
\epf

This construction can be generalized by using different matrices $Y_2$. Let $\pi_1,\pi_2,\ldots,\pi_{p+1}$ be permutations of 
the set $\{1,2,\ldots,p+1\}$ such that each permutation 
$\pi_j^{-1} \pi_k$, $j\ne k$, is fixed-point-free, i.e., 
$\pi_j^{-1} \pi_k(i)\ne i$ for all $i$. Then we can take $Y_2$ 
to be the matrix whose $(i,j)$th entry is $y'_{\pi_j(i),j}$. 
The row $i$ of $Y_2$ is 
$$
[~y'_{\pi_1(i),1}~y'_{\pi_2(i),2}~
\cdots~y'_{\pi_{p+1}(i),p+1}~].
$$
Since $\pi_j^{-1} \pi_k(i)\ne i$ whenever $j\ne k$, we have 
$\{\pi_j(i):j=1,2,\ldots,p+1\}=\{1,2,\ldots,p+1\}$. 
We infer that the row $i$ of $Y_2$ is orthogonal to $Y_1$.
Since $i$ is arbitrary, we have $Y_1\perp Y_2$.

There are many choices for the permutations 
$\pi_j$ that satisfy the condition stated above. For instance, 
we can choose any $(p+1)$-cycle $\sigma$ and set 
$\pi_j=\sigma^{j-1}$ for $j\in\{1,2,\ldots,p+1\}$. 
If $p=3$ and we choose $\sigma=(1432)$ then the matrix $Y_2$
is exactly the one shown in \eqref{eq:Y_1,Y_2}.
There are other choices as well. For instance, for $p=3$ we can take 
$\pi_1=id$, $\pi_2=(12)(34)$, $\pi_3=(13)(24)$ and 
$\pi_4=(14)(23)$, 
the elements of the Klein four-group. Then we obtain that
\bea \label{eq:KleinY_2}
Y_2=
\bma
y'_{11} & y'_{22} & y'_{33} & y'_{44} \\
y'_{21} & y'_{12} & y'_{43} & y'_{34} \\
y'_{31} & y'_{42} & y'_{13} & y'_{24} \\
y'_{41} & y'_{32} & y'_{23} & y'_{14} 
\ema.
\eea

Let us apply our construction in the case $n=7$. There are 
6 UOM-equivalence classes in $\cO(8,7)$. They are listed on 
Johnston's website and we denote them as $J_i$, $i=1,\ldots,6$. 
We list them here for the reader's convenience:

\begin{eqnarray*}
J_1 &=& [aaaaaaa, Abbbbbb, bABcccc, BBAdddd, cccABCD, CddBADC, dCDCDAB, DDCDCBA], \\
J_2 &=& [aaaaaaa, Abbbbbb, bABcccc, BcABddd, cdcABCD, CCddABC, dDDCDAB, DBCDCDA], \\
J_3 &=& [aaaaaaa, Abbbbbb, bAccBcc, BcABcdd, cdCAdBD, CCddACB, dDBDCAC, DBDCDDA], \\
J_4 &=& [aaaaaaa, Abbbbbb, bAcccBc, BcABdcd, cddABCC, CCCdAdB, dDBDCAD, DBDCDDA], \\
J_5 &=& [aaaaaaa, Abbbbbb, bAcccBc, BcAddcB, cddABCC, CCCBAdd, dDBCDAD, DBDDCDA], \\
J_6 &=& [aaaaaaa, Abbbbbb, bAccccB, BcAddBc, cddABCC, CCCBAdd, dDBCDAD, DBDDCDA]. \\
\end{eqnarray*}

For $k=1,2$ we set 
$$
X_k=\bma
a_k  & b_k  & c_k  \\
a'_k & e'_k & f_k  \\
d_k  & b'_k & f'_k \\
d'_k & e_k  & c'_k
\ema.
$$
For $Y_1$ we take the first matrix given 
in \eqref{eq:Y_1,Y_2} while for $Y_2$ we use four choices
for the permutations $\{\pi_i\}$: the above Klein four-group 
$K$, its coset $(1,2,3)K$, the cyclic group $C$ generated 
by $(1,2,3,4)$, and its coset $(1,2)C$. (The ordering of 
the $\pi_i$ is irelevant in these cases.)
By using our equivalence test (see Lemma \ref{le:eq-test}) we have verified that the four UOMs obtained in this way are 
equivalent to $J_1,J_2,J_6$ and $J_5$, respectively.
GenShift UOM is equivalent to $J_3$.

Our construction can be used to generate many UOMs of different 
sizes. Let us give a few examples. In these examples we use 
our construction $(X_1,X_2)\LRa Z$ by specifying $X_1$ to be an 
orthogonal matrix of size $m\times n$ while $X_2$ will always be obtained from $X_1$ by renaming the vector variables. Thus instead of $(X_1,X_2)\LRa Z$ we shall write just $X_1\LRa Z$. The auxilliary matrix $Y_1$ from our construction does not play an important role and we may consider it as being fixed. In this way we obtain a map $\cO(m,n)\to\cO(2m,m+n)$. Note that even when $X_1$ is an UOM, $Z$ is not necessarily an UOM.

For each positive odd integer $n$ we obtain an infinite sequence 
of maps:
$$
\cO(n+1,n)\to\cO(2n+2,2n+1)\to\cO(4n+4,4n+3)\to\cdots
$$
Each of the maps in this sequence preserves UOMs in the sense 
that the image of a UOM is again a UOM. This follows from 
Proposition \ref{pp:novuom}. By taking $n=1$, from the 
trivial UOM in $\cO(1)=\cO(2,1)$ we obtain UOMs in 
$\cO(4,3)$, $\cO(8,7)$, etc.

For each positive integer $n\equiv 2 \pmod{4}$ there is another infinite sequence of UOMs:
$$
X_n \LRa X_{2n+2} \LRa X_{4n+6} \LRa \cdots
$$
with $X_k\in\cO(k+2,k)$ for $k=n,2n+2,4n+6,\ldots$.
It is essential here that $X_n$ be chosen so that $\mu(x)=2$ 
for all entries $x$ in the first column, which implies that 
$\mu(x)=1$ for all other entries $x$ of $X_n$. 
It follows from the proof of \cite[Theorem 3.2]{kf06} that 
such $X_n$ exists when $n\equiv 2 \pmod{4}$.

For example, we can set $n=2$ and choose $X_2$ to be the third matrix in \eqref{eq:primer1}. Then by applying our construction, 
we obtain UOMs in $\cO(8,6)$, $\cO(16,14)$, etc.

\section{Construction of multiqubit PPT entangled states} \label{sec:pptes}

Let $\cS\subset\cH$ be an OPS. We say that the orthogonal projector $\r$ onto the subspace $\cS^\perp\subseteq\cH$ is 
{\em associated with} $\cS$. 
If $\Gamma$ is a partial transposition operator, then $\cS^\G$ 
is also an OPS and $\r^\G$ is the projector associated with it. This implies that $\r$ is a PPT state. 
In the case when $\cS$ is a UPB, then $\r$ is a 
(non-normalized) PPT entangled state (PPTES). This fact is 
also valid in arbitrary multipartite systems \cite{bdm99,DiV03}.

Even when $\cS$ is not a UPB, $\r$ may be a PPTES. Such OPS 
can be constructed from UOMs. The following lemma plays a crucial role in our construction.

\begin{lemma}
\label{le:uom-1}
Let $X=[x_{i,j}]\in\cO(m,n)$ be a UOM and let $Y$ be a submatrix of $X$ obtained by removing one of its rows. Then there exist 
only finitely many rows of vector variables which are orthogonal to $Y$. 
\end{lemma}
\bpf We may assume that the first row of $X$ has been removed.
Let $y:=[~y_1~y_2~\cdots~y_n~]$ be a row orthogonal to $Y$.
If $y'_1\notin\{x_{2,1},x_{3,1},\ldots,x_{m,1}\}$ then the row 
$[~x'_{1,1}~y_2~\cdots~y_n~]$ is orthogonal to $X$. As $X$ is a UOM, we have a contradiction. We conclude that 
$y_1\in\{x'_{2,1},x'_{3,1},\ldots,x'_{m,1}\}$. 
Similarly, we must have 
$y_j\in\{x'_{2,j},x'_{3,j},\ldots,x'_{m,j}\}$
for all $j$, and the lemma is proved.
\epf 

\bcr \label{cr:rangerho}
Let $X$ and $Y$ be as in the above lemma. Further, let $Z$ be the matrix obtained by appending to $Y$ all rows 
$u_1,u_2,\ldots,u_s$ orthogonal to $Y$. If $\a$ is a generic evaluation of $Z$, then the range of the projector $\r$ 
associated with the OPS of $\a(Y)$ contains only $s$ product 
vectors, namely $\a(u_1),\a(u_2),\ldots,\a(u_s)$.
\ecr

Let $X$ and $Y$ be as in Lemma \ref{le:uom-1}.
It is very easy to write a computer program which outputs all 
rows orthogonal to $Y$. Consequently, one can easily compute 
the product vectors in $\cR(\r)$ in Corollary \ref{cr:rangerho}.

If a projector is associated with a UPB, then it is a PPTES 
of the very special kind because its range contains no product vectors. Note that $s\ge 1$ in Corollary \ref{cr:rangerho}. 
If also $s \le 2^n-m$ then the projector $\r$ is a PPTES whose 
range contains $s\ge1$ product vectors. Hence, these PPTES 
are never equivalent (under local unitary transformations and qubit permutations) to those associated with the UPBs.

Let $X\in\cO(m,n)$, $m<2^n$. If $\a$ is a generic evaluation of $X$ then, by abuse of language, we say that the projector associated with the OPS of $\a(X)$ is also {\em associated 
with} $X$. 
Now assume that $X$ is an UOM. Then the projectors associated 
with $X$ are non-normalized PPTES and we say that they are the 
{\em primary PPTES} of $X$. 
Let us also introduce the {\em secondary PPTES} of $X$. These 
are the entangled projectors which are associated with the 
$(m-1)\times n$ submatrices $Y$ of $X$. Note that we have here 
singled out only the projectors associated with the $Y$s which 
are entangled. In general, a projector associated with $Y$ does 
not have to be entangled. For instance, when $n=3$ there are no 
secondary PPTES.

The secondary PPTES occur first for $n=4$.
In the following table, for each $m\in\T_4\setminus\{16\}$, we list all pairs $(\rank\r,s)$ where $\r$ is a secondary PPTES 
of some UOM $X\in\cO(m,4)$ and $s$ is the number of product 
vectors in $\cR(\r)$.

\begin{table}[ht]
$$
\begin{array}{ccc}
m & \rank\r & s \\
\hline \\
6 & 11 & 10 \\
7 & 10 & 3,6,7,8 \\
8 & 9 & 1,2,3,4,6 \\
9 & 8 & 1,2,3,4,6 \\
10 & 7 & 1,2,4 \\
12 & 5 & 1
\end{array}
$$
\caption{Secondary PPTES of four qubits}
\label{tab:secondary}
\end{table}

We point out that the rank-5 PPTES for $m=12$ (see the last line of Table \ref{tab:secondary}) support our conjecture in \cite[Conjecture 10]{cd17}. In the case $m=6$ there is only one 
UOM up to equivalence. We may assume that this is the matrix 
$Y$ in \eqref{eq:xy}. If we delete the first or fourth row of 
$Y$ then the projector $\r$ is separable. In the other four 
cases it is entangled of rank 11 and has 10 product vectors in 
its range.

Let us give yet another example. Consider the following UOM in $\cO(7,4)$ 
\begin{equation} \label{eq:primer4q} 
X=\left[ \begin{array}{cccc}
a  & c  & e  & g  \\
a  & c' & f  & h  \\
a  & d  & f' & g' \\
a' & c  & e  & i  \\
a' & d  & f  & i' \\
b  & c' & f' & g  \\
b' & d' & e' & h'
\end{array} \right]. 
\end{equation}
Drop the first row, $u$, from $X$ to get $Y$. Apart from $u$ there are only two other rows orthogonal to $Y$, namely 
$v:=[~a~c~f~h~]$ and $w:=[~a~d~f~h'~]$. Since in this case 
$v$ and $w$ are orthogonal, if we append $v$ and $w$ to $Y$ we 
obtain an UOM $Z\in\cO(8,4)$. Let $\a$ be a generic evaluation of $Z$ and $\cS$ the OPS of $\a(Y)$. If $\r$ is the projector 
associated with $\cS$, then $\cR(\r)$ contains only three product vectors, namely $\a(u)$, $\a(v)$ and $\a(w)$. As $\r$ has rank 10, it follows that $\r$ is entangled. Hence, 
$\r$ is a PPTES.

We can also construct PPTES by dropping more than one row 
from a UOM. We give an example. Let $X$ be the $10\times5$ 
UOM with rows:
$$
[baBCc,cbaBC,CcbaB,BCcba,AAAAA,CaCcb,aACbb,cBbBa,CABBa,caCbC].
$$
Let $Y\in\cO(8,5)$ be the matrix obtained by dropping the last two rows of $X$. Then there are exactly 6 rows orthogonal to $Y$,
namely the two last rows of $X$ and the following four:
$aABBc,~caAcA,~cabbA,~CAaBb$. 
The projector associated with $Y$ has rank 24 and has only 6 product vectors in the range. Hence it is a PPTES.

The construction of PPTES described above for multiqubit 
systems can be generalized to arbitrary finite-dimensional 
quantum systems, see Proposition \ref{pp:kon} in 
Sec. \ref{app:B}.

\section{A partial order in $\cM(m,n)$} \label{sec:maximal}

Let $X\in\cM(m,n)$ and let $x,y$ be two independent vector 
variables which occur in the same column of $X$, say column $j$.
Denote by $Y$ the matrix obtained from $X$ by replacing all occurrencies of $y$ and $y'$ in $X$ as follows: $y$ by $x$ and 
$y'$ by $x'$ or $y$ by $x'$ and $y'$ by $x$. Then we shall write $Y\prec X$. Note that we have $\nu_j(Y)=\nu_j(X)-1$, i.e., the 
number of independent variables of $Y$ is one less than that 
of $X$.

The relation ``$\prec$'' induces the partial order ``$\le$'' on 
$\cM(m,n)$, known as the transitive closure of ``$\prec$''. 
Explicitly, for two matrices $X,Y\in\cM(m,n)$, we say that 
$Y\le X$ if there exists a finite chain 

\begin{equation} \label{eq:chain}
Y=Z_0\prec Z_1\prec \cdots \prec Z_k =X,\quad k\ge 0. 
\end{equation}
Further, we write $Y<X$ if $Y\le X$ and $Y\ne X$. 

Assume that $Y\prec X$. Then it is easy to see that 
$X\in\cO(m,n)$ implies that $Y\in\cO(m,n)$. 
However, the converse is false.
Further, if $X$ is a UOM then $Y$ is not necessarily a UOM. 
For instance, if we replace $f$ and $f'$ 
with $e$ and $e'$ in the UOM $X\in\cO(4,3)$ in \eqref{eq:xy} 
then the resulting matrix is not a UOM. 

Consequently, if in the chain \eqref{eq:chain} we have 
$X\in\cO(m,n)$ then all $Z_i\in\cO(m,n)$.

If $X$ and $Y$ in \eqref{eq:chain} are UOM then so are all 
the $Z_i$. This follows from the following lemma.

\bl \label{le:chain}
If $Y\prec X\in\cO(m,n)$ and $Y$ is a UOM then $X$ is a UOM.
\el
\bpf
By the hypothesis, we may assume that $Y$ is obtained from $X$ by identifying two independent variables $a$ and $b$ in the first column of $X$. Say, we replaced each occurrence of $b$ 
with $a$ (and $b'$ with $a'$).

Suppose that there exists a row of vector variables 
$u:=[~u_1~u_2~\cdots~u_n~]$ which is orthogonal to $X$. 
Since $Y$ is a UOM, we may assume that $u$ is not orthogonal to 
the first row $y:=[~y_1~y_2~\cdots~y_n~]$ of $Y$. Hence, the first row $x:=[~x_1~x_2~\cdots~x_n~]$ of $X$ is not equal to 
$y$. It follows that $x_1\in\{b,b'\}$ and we may assume that 
$x_1=b$, and so $y_1=a$. It is now easy to verify that the 
row $v:=[~a'~u_2~\cdots~u_n~]$ is orthogonal to $Y$, which 
gives a contradiction.
Thus we have shown that $X$ has no 1-extensions, i.e., it
is a UOM.
\epf

We say that a UOM $X\in\cO(m,n)$ is {\em maximal} if there is no UOM $Y\in\cO(m,n)$ such that $X<Y$. Similarly, we say that a UOM $X\in\cO(m,n)$ is {\em minimal} if there is no UOM 
$Y\in\cO(m,n)$ such that $Y<X$. Further we say that an UOM is {\em isolated} if it is both minimal and maximal. 
In order to prove that a UOM $X$ is maximal, by the above lemma, 
it suffices to verify that there is no UOM $Y\in\cO(m,n)$ such that $X\prec Y$. Similarly, in order to prove that a UOM $X$ is minimal, it suffices to verify that there is no UOM $Y\in\cO(m,n)$ such that $Y\prec X$. 

These definitions extend naturally to equivalence classes of orthogonal matrices and UOMs. For instance, if $\cX$ and $\cY$ are two equivalence classes of matrices in $\cO(m,n)$ and 
$Y\le X$ for some $X\in\cX$ and some $Y\in\cY$, then we shall write $\cY\le\cX$. We say that an equivalence class of UOMs $\cX\subseteq\cO(m,n)$ is {\em maximal} if there is no equivalence class of UOMs $\cY\subseteq\cO(m,n)$ such that 
$\cX<\cY$, etc. 

Finally, we say that a UOM $X$ and its equivalence class $[X]$  are {\em irreducible} if all $\nu_j(X)>1$. Otherwise, we say that $X$ and $[X]$ are {\em reducible}. Any reducible UOM $X$ 
is equivalent to one of the form 
$\left[\begin{array}{c}a\\a'\end{array}\right]\models(U,V)$, 
where $U$ and $V$ are UOMs. Thus all reducible UOMs are 
decomposable.

\bl \label{le:redUOM}
Let $X$ be a decomposable UOM, say $X=A\models(B_1,\ldots,B_r)$.
Then 

(i) $X$ is maximal if and only if $A$ and all $B_i$ are maximal and no two of the blocks $B_i$ have a vector variable in common; 

(ii) if $X$ is not maximal then there exists a UOM $Y$ 
such that $X\prec Y$ and $Y$ is obtained from $X$ by modifying 
a single column in either $A$ or just one of the blocks $B_i$.
\el
\bpf
(i) Assume $X$ is maximal. By Proposition \ref{pp:konstr}, $A$ and all $B_i$ are UOMs. Assume that some $B_i$ is not maximal. Then $B_i\prec U$ for some UOM $U$. Hence, $B_i$ and $U$ differ only in one column and $U$ has one new independent variable, 
say $u$. We may assume that $u$ is independent from the variables which occur in $X$. Let $Y$ be the UOM obtained from $X$ by replacing $B_i$ with $U$. Then $X\prec Y$ and we have a contradiction since $X$ is maximal. We conclude that each $B_i$ must be maximal. Similarly, one can show that $A$ has to be maximal. It remains to consider the case where $A$ and all the 
$B_i$ are maximal. Assume now that one of the vector variables, 
say $u$, occurs in $B_i$ and $B_j$ (necessarily in the same column). We choose a new vector variable $v$ which does not 
occur in $X$. Let $Y$ be the matrix we obtain by replacing all occurrences of $u$ and $u'$ in $B_j$ with $v$ and $v'$, respectively. Then $X\prec Y$ and we have again a contradiction.
Hence our assumption must be false, i.e., no vector variable 
can occur in two blocks $B_i$ and $B_j$. This completes the proof of the ``if'' part.

The ``only if'' part follows immediately from the definition of 
``$\prec$''.

(ii) This follows from the proof of (i) by observing that the 
matrix $Y$ constructed there differs from $X$ in a single 
column in either $A$ or just one of the blocks $B_i$.
\epf

For example the UOMs
\bea
\label{eq:prvi-model}
&& \left[\begin{array}{c}a\\a'\end{array}\right]\models
\left(
\left[\begin{array}{ccc}b&c&d\\b'&e&f'\\g'&c'&f\\g&e'&d' \end{array}\right],
\left[\begin{array}{ccc}u&v&w\\u'&x&y'\\z'&v'&y\\z&x'&w' \end{array}\right]
\right), \\
\label{eq:drugi-model}
&& \left[\begin{array}{ccc}a&b&c\\a'&d&e'\\f'&b'&e\\f&d'&c' \end{array}\right]\models
\left(
\left[\begin{array}{c}u\\u'\end{array}\right],
\left[\begin{array}{c}v\\v'\end{array}\right],
\left[\begin{array}{c}w\\w'\end{array}\right],
\left[\begin{array}{c}x\\x'\end{array}\right]
\right) 
\eea
are maximal in $\cO(8,4)$. The first one is reducible and the 
second one irreducible.

We remark that all UOM-equivalence classes can be derived from the knowledge of the maximal classes. To explain how this is 
done we need to define some subsets of $\cO(m,n)$ which we 
call ``levels''. The {\em level $l$} of $\cO(m,n)$ is the 
set $\{X\in\cO(m,n):\nu(X)=l\}$. (The function $\nu$ is 
defined in Sec. \ref{sec:pre}.) For convenience, we say that 
$X\in\cO(m,n)$ lies on level $l$ if $\nu(X)=l$. 
Note that if $X$ lies on level $l$ then the whole equivalence class $[X]$ also lies on the same level.

{\em Sketch of the algorithm.}
We first choose a set of representatives, say $R$, of the set of maximal classes. Next, from $R$ we extract the matrices which lie on the highest level, say $l$. To each $X$ in the selected set, we apply the following procedure. We construct all matrices $Y$ on level $l-1$ such that $Y\prec X$. There are only finitely many such $Y$s. We discard those $Y$ which are not UOM, say by using Lemma \ref{le:DiV03}.
If no $Y$s are left, then the class $[X]$ is minimal.
After performing this procedure on all maximal $X$ on level $l$,
we test the $Y$s for equivalence. If two $Y$s are equivalent, 
we remove one of them. We repeat this step until the remaining $Y$s become pairwise non-equivalent. Note that none of these 
$Y$s is maximal. Next, we enlarge the set of $Y$s by adding the matrices in $R$ which lie on level $l-1$ (if any). Then we repeat the same procedure on the new set of $Y$s that we constructed. We carry out this process to its end where 
no new matrices show up and all matrices of $R$ have been 
used up. The computation is admittedly tedious but it could be programmed to perform all steps on the computer.
One of the hard steps is the test of equivalence.

\section*{Acknowledgements}

LC was supported by Beijing Natural Science Foundation (4173076), the NNSF of China (Grant No. 11501024), and the Fundamental Research Funds for the Central Universities (Grant Nos. KG12001101, ZG216S1760 and ZG226S17J6). 
The second author was supported in part by the National Sciences and Engineering Research Council (NSERC) of Canada Discovery 
Grant 5285.

\section{Appendix A} \label{app:A}

\begin{table}[h]
\caption{UOMs of new sizes}
\label{tab:sizes}
$
\begin{array}{ll}
{\rm Size} & \text{\rm UOM as a list of rows 
$(A=a',~B=b',\ldots)$} \\
\hline \\
11\times 5 &
[cbaBC, CcbaB, BCcba, AAAAA, aBCBb, baBbc, CCcBa, aCCbC, cacBc, 
 ccbba, BcBba] \\
10\times 6 & 
[aaaaaa, Abbcbb, bBcACd, bcCBAB, BABddC, cBCbDA, CCADBD, \\
&  BccCAc, bAcCcd, bBcdAD] \\
11\times 6 & 
[aAbbbb, Abcccc, ABdddd, cCBADC, CdCBAD, CDDCBA, \\
& adaaaB, caDaCb, CDBDCa, ccADCB, aDdBcc] \\
13\times 6 &
[aAbbbb, Abcccc, ABdddd, bcADCB, bCBADC, BdCBAD, BDDCBA, \\ 
& acaBaa, aCBaaa, aCbcaB, acBbca, bCbCBD, aabbbb], \\
& [aAbbbb, Abcccc, ABdddd, bcADCB, bCBADC, BdCBAD, BDDCBA, \\
& aDaaBa, BaDCbd, BdBCaD, adbdBd, BDCBbD, BdCDBd] \\
13\times 7 &
[baBCDdc, cbaBCDd, dcbaBCD, DdcbaBC, CDdcbaB, BCDdcba, \\
& AAAAAAA, aBCCCcC, aBDDdcc, BadBccb, aAdCcbd, aDDDccC, cadbdcc], \\ 
& [baBCDdc, cbaBCDd, dcbaBCD, DdcbaBC, CDdcbaB, BCDdcba, \\
& AAAAAAA, DBCDaDb, dBaDdcb, adADAcC, DcaBddb, dAADaAd, DdabABC] \\
14\times 7 &
[baBCDdc, cbaBCDd, dcbaBCD, DdcbaBC, CDdcbaB, BCDdcba, AAAAAAA, \\
& aBCCCcC, aBDDdcc, BadBccb, aAdCcbd, aDDDccC, cadbdcc] \\
15\times 7 &
[baBCDdc, cbaBCDd, dcbaBCD, DdcbaBC, CDdcbaB, BCDdcba, \\
& AAAAAAA, DDaDddb, aBCDbDb, adCDddd, aBBDBDc, dcbABaD, \\
& dcbaBcD, dCbaBBD, DDADBaC] \\
19\times 7 & 
[baBCDdc, cbaBCDd, dcbaBCD, DdcbaBC, CDdcbaB, BCDdcba, \\
& AAAAAAA, DDaDddb, aBCDbDb, acAcBcc, adCACdC, aCCDBDD, \\
& ddbabdA, CcaABDc, BdaACdc, ddBaCdC, dCbaBdA, bdaAddc, DcAaBbC] \\
13\times 8 
& [aaaaaaaa, bbbAbbbb, cccbcABc, CdBcDdAe, BdAdCDdD, \\
& BDDBeeeA, CDAdBEdD, ABDDdcEC, dCdDABCE, DACCECDB, \\
& ABdCCccd, BDdcCecA, BdAdcDdC] \\
14\times 8 
& [aaaaaaaa, bbbAbbbb, cccbcABc, CdBcDdAe, BdAdCDdD, \\
&  BDDBeeeA, CDAdBEdD, ABDDdcEC, dCdDABCE, DACCECDB,  \\
&  ABdCCccd, DDdcCeAB, BdAdcDdC, DBdcCAad] \\
15\times 8 
& [aaaaaaaa, bbbAbbbb, cccbcABc, CdBcDdAe, BdAdCDdD, \\
&  BDDBeeeA, CDAdBEdD, ABDDdcEC, dCdDABCE, DACCECDB, \\
&  ABdCCccd, DDdcCeAB, BdAdcDdC, DAdcCBad, BAdcCbad] \\
17\times 8 & 
[bbbAebbd, dbbBEbDA, DDdAbcbD, DAaBECab, DabAECbd, AdbdedbD, \\
& bdCdeDAD, aaaaaaaa, cccbcABc, ABdBdccd, CdBcDdAe, BdAdCDdD,\\
& BDDBeeeA, CDAdBEdD, ABDDdcEC, dCdDABCE, DACCECDB] \\
18\times 8 & 
[DbACEcbd, ebbAebbd, dbbBEbDA, DDdAbcbD, DAaBECab, \\
& DabAECbd, AdbdedbD, bdCdeDAD, aaaaaaaa, cccbcABc, \\
& ABdBdccd, CdBcDdAe, BdAdCDdD, EDDBeeeA, CDAdBEdD, \\ 
& ABDDdcEC, dCdDABCE, DACCECDB] \\
19\times 8 & 
[DbCCEcAd, DbCAEcad, ebbAebbd, dbbBEbDA, DDdAbcbD, DAaBECab, \\
& DabAECbd, AdbdedbD, bdCdeDAD, aaaaaaaa, cccbcABc, ABdBdccd, \\
& CdBcDdAe, BdAdCDdD, EDDBeeeA, CDAdBEdD, ABDDdcEC, dCdDABCE, \\ 
& DACCECDB] \\
\end{array}
$
\end{table}

\newpage

\section{Appendix B} \label{app:B}

Corollary \ref{cr:rangerho} shows that if $\cU$ is a UPB in 
a multiqubit system and $\cU'\subset\cU$, $|\cU'|=|\cU|-1$,
then the subspace ${\cU'}^\perp$ contains only finitely 
many product vectors. We show here that this fact is valid 
in arbitrary finite-dimensional quantum systems.
Thus in this appendix we drop the condition that 
$\dim\cH_i=2$ for all $i$, and we set $d_i=\dim\cH_i$ 
for $i=1,2,\ldots,n$. Then $d:=\prod d_i$ is the dimension 
of $\cH$.

Moreover, the fact mentioned above remains valid when we replace 
UPBs with generalized UPBs. They are defined as follows.

A {\em generalized UPB} (abbreviated as gUPB) is a linearly independent set $\cU\subset\cH$ of unit product vectors such that $\cU^\perp$ contains no product vector.

\bpp \label{pp:kon}
Let $\cU$ be a gUPB of cardinality $m$ in $\cH$ and let 
$\cU'$ be a subset of $\cU$ of cardinality $m-1$. 
Then the subspace ${\cU'}^\perp$ contains only finitely 
many product vectors (up to scalar multiples). 
\epp
\bpf
The Segre variety $\cS:=P(\cH_1)\times \cdots \times P(\cH_n)$ 
is embedded in the complex projective space $P(\cH)$ as 
a closed subvariety. The projective linear subspaces 
$\Lambda:=P(\cU^\perp)$ and $P({\cU'}^\perp)$ of $P(\cH)$ 
have dimensions $d-m-1$ and $d-m$, respectively. 
Set $X:=P({\cU'}^\perp)\cap\cS$, a closed subvariety of 
$P({\cU'}^\perp)$, and let $k$ be its dimension.

Assume that $k\ge1$. By applying \cite[Proposition 11.4]{ha92} 
to the projective space $P({\cU'}^\perp)$ of dimension $d-m$, 
its subvariety $X$ of dimension $k$, and the linear subspace $\Lambda$ of dimension $d-m-1$, we deduce that $\Lambda$ must 
intersect $X$. As $X\subseteq\cS$, $\Lambda$ also intersects 
$\cS$. This contradicts the hypothesis that $\cU$ is a gUPB. 
We conclude that $k<1$, i.e., $X$ must be a finite set 
(possibly empty).
\epf

As an example let us consider the case of two qutrits: 
$n=2$, $d_1=d_2=3$, $d=9$. Let $\cU$ be the {\em Pyramid} UPB 
in $\cH$ (see e.g., \cite{DiV03}). It consists of five product 
states $\psi_i=v_i\ox v_{2i \pmod{5}}$, $i=0,\ldots,4$, where 
$$
v_i=N\left(\cos\frac{2\pi i}{5},\sin\frac{2\pi i}{5},h\right), 
\quad i=0,\ldots,4,
$$
with $h=\frac{1}{2}\sqrt{1+\sqrt{5}}$ and 
$N=2/\sqrt{5+\sqrt{5}}$. Let $\cU'$ be the OPS consisting of 
the four states $\psi_i$, $i=0,\ldots,3$. Denote by $\r$ the 
projector associated with $\cU'$. One can verify that 
the range of $\r$ contains exactly six product states. 
To write down these product states, we denote by $u_{i,j}$, 
$i\ne j$, the unit vector orthogonal to $v_i$ and $v_j$. 
Then the six product vectors in the range of $\r$ are 
$\psi_4=v_4\ox v_3$, $u_{2,4}\ox v_3$ and the four mutually orthogonal states 
$v_0\ox u_{0,2}$, $v_3\ox u_{1,4}$, 
$u_{1,3}\ox v_2$, $u_{0,2}\ox v_4$. 
None of the last four states is orthogonal to $\psi_4$ or 
$u_{2,4}\ox v_3$. By using a computer we have verified that 
$\r$ and the density matrices of the six unit product 
vectors in the range of $\r$ are linearly independent. 
It follows that $\r$ is not separable. To summarize, $\r$ is a (non-normalized) PPTES of rank 5 whose range contains exactly 6 product vectors. 

While in the multiqubit case we can easily find all product 
vectors in the range of $\r$, this example shows that in general 
this task is not easy.

\end{document}